\documentstyle[12pt]{article}
\begin{document}
\begin{titlepage}
\begin{flushright}
INFNCA-TH-94-30
\end{flushright}
\vskip 10mm
\begin{center}
\par \vskip 10mm
{\large  \bf      On the Plasmon-like Excitations of Atomic Clusters} \\
\vskip 2mm
{\large  \bf       with Application to Fullerene}
\end{center}
\par \vskip 2mm
\begin{center}
               B.\ Carazza
\end{center}
\par
\begin{center}
{\it Dipartimento di Fisica dell' Universit\`a,
   viale delle Scienze, I43100 Parma, Italy}\\
      {\it INFN Sezione di Cagliari, Italy}\\
\end{center}
\par \vskip 2mm
\begin{center} {\bf  Abstract} \end{center}
 \vskip 1mm
\begin{quote}
We present here a modified hydrodynamical model which can be
used to investigate in a easy way the collective electronic
excitations of spherical atomic clusters. The model is then
applied to the case of the giant-dipole plasmon mode of the
free fullerene molecule.
\end{quote}
\vskip .5 true cm
PACS numbers: 71.45.Gm, 36.40.+d
\vspace*{\fill}
\end{titlepage}
\setlength{\baselineskip}{25pt}
      Collective electronic excitations of small metallic
      particles or atomic clusters like fullerites
      may be and in fact are currently [1-6]
      %\cite{Sel89,Bre89,Ber91,Gen91,Bul92,Soh92}
      of great interest with the goal to bridge the gap
      between  our description of atom and solid.
      A part the polarization propagator response method \cite{Mahbo}
      within RPA  \cite{Pin52,Ber91}
      a good tool to investigate theoretically
      the plasmon-like collective behaviour of the nearly
      free or valence electrons in such systems is certainly
      the hydrodynamical model which date back to pioneering
      work by Bloch \cite{Blo33,Blo34}. It was early used  \cite{Jen37}
       eventually in refined version, as for instance in
      the investigation of photoabsorption of a Thomas-Fermi
      atom \cite{Bal73}. More recently such an approach was successfully
      used investigating the excitation of plasmons by energetic
      electrons in small spheres \cite{Cro68} and photon
      bombardment of a plasma slab \cite{Wil67}.
      Since then the model has been refined, \cite{Marbo}
      and now leads to very sophisticated equations
      for the density fluctuations around the equilibrium
      distribution of a inhomogeneous electron gas \cite{Thebo}.
      In spite of that, this approach seems not so widely
      used, probably because the necessary equations are
      very cumbersome to handle unless the equilibrium
      density distribution can be considered constant.
      Thus I propose in this note a way to simplify these
      difficulties in the case of a problem with spherical symmetry,
      which will be presented within the more simple
      Bloch's original scheme.
      It  consists on the idea
      to consider the equilibrium electronic density  $\rho_0(r)$
      which is isotropic for a problem with spherical symmetry,
      to  assume  constant values  $\rho_0^{(i)}$ within each shell
      of a given
      thickness $\Delta_i$ by which  the space around the origin
      can be subdivided.
      I will also consider the quantity above to be
      zero outside a certain distance $R$ form the origin.
      The case of $\rho_0(r)$ being continuous is of course recovered
      if the shell's thickness goes to zero and
      $R$, if necessary, to infinity. But all that amounts to
      present an approximate method to solve the continuous
      case, with no less difficulties than before if you like
      to approach more and more the limit. Instead, also
      in view of the validity of the approximation leading
      to $\rho_0(r)$ both
      the set of values $\rho_0^{(i)}$
      as well $R$ and the $\Delta_i$ will be considered as
      parameters referring to a model equilibrium density.
      The net result will be a kind of multi-shell or onion
      model.
      \par
      The starting point in the Bloch's approach is a Lagrangian
      ${\cal L}_2$.
      It  comes as the second order (quadratic) contribution
      in the development around equilibrium of the full
      Lagrangian ${\cal L}$ for
      an irrotational fluid of electrons in which only the Coulomb
      interaction with the positive background, the Coulomb interaction
      among the electrons and
      the kinetic energy term are considered. The first variation
      of ${\cal L}$ in the static case leads to the Thomas-Fermi
      equation.
      The equations of motion ( written in atomic units to
      be used in the following ) result to be:
      \begin{equation}\label{1}
       \dot{\phi}(\mbox{\boldmath $r$},t)={C[\rho_0(r)] \over \rho_0(r)}
       \,n(\mbox{\boldmath $r$},t)+\int {n(\mbox{\boldmath $r$}',t) \over
       |\mbox{\boldmath $r$}-\mbox{\boldmath $r$}'|}
      \,d\mbox{\boldmath $r$}'       \qquad \qquad
      \dot{n}(\mbox{\boldmath $r$},t)=\mbox{\boldmath $\nabla$}\cdot
       [\rho_0(r) \mbox{\boldmath $\nabla$}
      \phi(\mbox{\boldmath $r$},t)]
      \end{equation}
      where $n(\mbox{\boldmath $r$},t)$ and
      $\phi(\mbox{\boldmath $r$},t)$ are perturbations in the
      electronic density and velocity potential , respectively,
      in the electron gas.
       As already indicated ${\rho_0(r)}$ is the electronic
      density in the undisturbed state so that
       $\rho_0(r) \mbox{\boldmath $\nabla$}
       \phi(\mbox{\boldmath $r$},t)$ represents at this order
      of development the current density of the fluctuations.
      $C[\rho_0(r)]$ is the quantity $\pi^2/3 (3/\pi)^{2/3}
      \rho_0(r)^{2/3}$.
      Assuming:
     \begin{equation}\label{2}
      \phi(\mbox{\boldmath $r$},t) = f(\mbox{\boldmath $r$}) \left
      \{ \begin{array}{r}
        -\cos \omega t \\ \sin \omega t \end{array} \right. \quad \quad
      n(\mbox{\boldmath $r$},t) =  g(\mbox{\boldmath $r$})
       \left \{ \begin{array}{r}
       \sin \omega t \\ \cos \omega t \end{array} \right.
      \end{equation}
      we have :
      \begin{eqnarray}\label{3}
       \omega f(\mbox{\boldmath $r$}) &=& {C[\rho_0(r)] \over \rho_0(r)}
      \,g(\mbox{\boldmath $r$})+\int
        {g(\mbox{\boldmath $r$}') \over |\mbox{\boldmath $r$}-
      \mbox{\boldmath $r$}'|} \,d\mbox{\boldmath $r$}' \nonumber \\
       \omega g(\mbox{\boldmath $r$}) &=& -\mbox{\boldmath $\nabla$} \cdot
       [\rho_0(r) \mbox{\boldmath $\nabla$}  f(\mbox{\boldmath $r$})] \, .
      \end{eqnarray}
      In the case that $\rho_0(r)$ assumes a constant value
      $\overline\rho_0$
      and using both the Eq.~(\ref{3}) one see that $g(\mbox{\boldmath $r$})$
      satisfies the equation :
      \begin{equation}\label{4}
      \triangle g(\mbox{\boldmath $r$})+ {(\omega^2 -4 \pi
      \overline\rho_0) \over C[\overline\rho_0]} \,g(\mbox{\boldmath $r$})=0
      \end{equation}
      which is easy solvable after separation of the angular
      dependence. In the present case it is convenient to
      do so using the real, normalized spherical harmonics
      ${\cal Y}_{m,l}^s(\Omega)$ \cite{Morbo} where $s$ means
      $e$ (even) or $o$ (odd),
      in order to easy obtain $f(\mbox{\boldmath $r$})$ and
      $g(\mbox{\boldmath $r$})$
       as real functions as      it should be.
      Let $F= (\omega^2 -4 \pi \overline\rho_0) /
      C[\overline\rho_0]$  and $\mu =\sqrt{|F|} $.
      The general solution of Eq.~(\ref{4}) is of the type:
      \begin{equation}\label{5}
      g_{l,m}^s(\mbox{\boldmath $r$})=[a \,A_l(\mu r)+b \,B_l(\mu r)]
      \,{\cal Y}_{m,l}^s(\Omega)
      \end{equation}
      where $a$, $b$ are numerical coefficient and  $A_l(\mu r)$,
      $B_l(\mu r)$ indicate the couple $j_l(\mu r)$ and $\eta_l(\mu r)$
      of independent Bessel functions of order $l$
      when $F$ is positive or their
      analytical extension      in terms of hyperbolic
      sinus and cosinus otherwise.
      Consider now a isotropic equilibrium density
      distribution which is zero outside the sphere of
      radius $R$ and let $R$ be subdivided for simplicity in $N$
      equal intervals of length $ \Delta= R/N$.
      The sphere is thus subdivided in $N$ shells, the $i^{th}$
      among them having $r_{i-1}$ and  $r_i$  as internal
      and external radius respectively with
      $r_i=i \Delta \quad (i=1, N)$  and let the quantity $\rho_0^{(i)}$
      be the constant value of the equilibrium density
      within the $i^{th}$ shell.
      The solutions for $g(\mbox{\boldmath $r$})$  will be:
      \begin{equation}\label{6}
      g_{l,m}^s(\mbox{\boldmath $r$})=\displaystyle
      \,\sum_{i=1}^N[a_{l,m}^{(i) s} \,
      A_l(\mu_i r)+ b_{l,m}^{(i) s} \,B_l(\mu_i r)] \,
      {\cal Y}_{m,l}^s(\Omega)  \,     \chi^{(i)}(r)
      \end{equation}
      where $\chi^{(i)}(r)$ is the step function which is equal to one
      inside the $i^{th}$ shell and zero otherwise, and the
      meaning of the quantities indexed with $i$ is obvious.
      The corresponding solution for the velocity potential
      is obtained from Eq.~(\ref{6}) using the first equation
      in Eq.~(\ref{3}).
      It is required that both $f(\mbox{\boldmath $r$)}$ and
      $g(\mbox{\boldmath $r$})$
      be regular at the origin and as usual that the
      normal component      of the fluctuation current
      density at the surface $r = R$ be zero. Besides
      these boundary conditions  also the  matching conditions at the
      separating surfaces of the shells are required.
      For that we have at disposal the values of the fields
      $\phi$ and $n$ and of
      their derivatives at such places.
      As $\rho_0(r)$ is considered discontinuous I see no need
      to impose continuity conditions on the fluctuation
      density. So we can concentrate on $\phi$ requiring the
      continuity of the velocity potential and of the
      fluctuation      density current $\rho_0(r)
      \mbox{\boldmath $\nabla$} \phi(\mbox{\boldmath $r$},t)$.
      The regularity of the fields at $r = 0$ requires
      that the coefficient of the non regular Bessel function
      referring to the first shell around the origin be zero.
      The boundary condition at $r = R$ and the $2N-2$
      conditions at the separation walls then give rise
      to $2N-1$ homogeneous linear equations for the remaining
      $2N-1$ unknown coefficients
      and we are left with an eigenvalue problem.
      Let $\omega_n$ be an eigenvalue and $f_n(\mbox{\boldmath $r$)}$
       \ $g_n(\mbox{\boldmath $r$)}$ the
      couple of corresponding solutions for $f(\mbox{\boldmath $r$)}$
      and $g(\mbox{\boldmath $r$)}$.
      It is easy to show \cite{Note} that thanks to the conditions
      imposed the integral
      of the product      $f_m(\mbox{\boldmath $r$)}
      g_n(\mbox{\boldmath $r$)}$ over the volume
      $|\mbox{\boldmath $r$}| < R$ is zero       unless $m=n$,
      that is the result which holds when $\rho_0(r)$
      is continuous \cite{Blo33,Bal73} is recovered.
      The solution will then be normalized with the
      following choice :
      \begin{equation}\label{7}
      \int f_n(\mbox{\boldmath $r$)} \,
      g_m(\mbox{\boldmath $r$)} d\mbox{\boldmath $r$}=
      \delta_{nm} \,\omega_n   \, .
      \end{equation}
      I notice that a choice  of an
      orthogonal set of harmonic functions for the
      angular part was already made, so the degeneracy
      connected with the spherical symmetry does not
      cause trouble in this respect.
      \par
      To illustrate the method outlined above I applied the onion
      model to the collective electronic excitations of fullerene,
      since this recently produced macromolecule is
      the subject of many investigation both
      on experimental and theoretical side.
      The $C_{60}$ is considered with good approximation \cite{Sir92}
      as a neutral system composed of nearly-free electrons
      with the background positive charge uniformly
      distributed over the surface of a sphere of radius $r_b$.
      As a starting point  the Thomas-Fermi
      equilibrium distribution $\rho_{0TF}(r)$ of the valence
      electrons, whose number was fixed to be 240
      following the Ref.~\cite{Ash91}, was computed.
      Then the following quantities were considered for
      each shell:
      \begin{equation}\label{8}
      \overline\rho_{0i}={1 \over V_i} \int \rho_{0TF}(r) \,dV_i
      \end{equation}
      where $V_i$ is the volume of the $i^{th}$  shell.
      The required constant quantities $\rho_0^{(i)}$ were then
      obtained as $\overline\rho_{0i}$ time a normalization factor
      with the aim to ensure that the total number of electrons
      contained in the volume $|r| < R$ be exactly 240.
      The eigenvalue problem was solved numerically for
      three different values of the fullerene radius
      $r_b$ in the only case for which  $l=1$.
      The interval of explored frequencies ranges from 7
      to 32 eV, in order to explore
      the discovered high-energy giant-dipole plasmon mode
      of the free molecule \cite{Her92}.
      We know that it exists a lower limit of distance
      given by $1/Z$ where $Z$ is the charge, for the
      applicability of Thomas-Fermi equation
      near a charged nucleus
      (in the present case the carbon nucleus screened by the
      two inner s-electrons) \cite{Lanbo}, and therefore
      the detailed behaviour of the resulting equilibrium electronic
      density at shorter distance can be ignored.
      We also know that the  ground state
      electronic distribution of a Thomas-Fermi atom
      has a tail going wrong \cite{Mesbo},
      i$.$e.\ it goes too much slowly to zero at infinity,
      and as a crude remedy we can cut off the obtained
      $\rho_{0TF}(r)$ at some finite radius.
      Taking into account the considerations above the values
      $\Delta = 0.2$
      and $R = 14.2 $ were adopted as sensible.
      The last figure is about two
       time the value of $C_{60}$ radius i$.$e.\ 3.5 \AA \cite{Fisbo}
      and so the equilibrium distribution ends at fairly
      the same radial distance from the positive charged
      surface.
      After the eigenvalues and the related eigenfunctions
      are known the next step is to build up the Hamiltonian
      which using these normal mode will look of course like
      a collection of non interacting harmonic oscillator
      hamiltonians. Let us restrict to the
      space of functions whose angular dependance is that
      of the usual $l=1, m=0$ harmonic. Indicating with
      $s_k(\mbox{\boldmath $r$)}$ and $w_k(\mbox{\boldmath $r$)}$
      the $k^{th}$ solution  with the above angular part for
      $f(\mbox{\boldmath $r$)}$ and $g(\mbox{\boldmath $r$)}$
      respectively and expanding  with such
      a basis the fields $\phi$ and $n$ it is easy to build
      up the contribution to the total Hamiltonian as coming
      from the subspace considered. It is also easy to write
      down the dipole interaction term when an external electric
      field $E$ directed along the polar axis is considered.
      At last one see that the oscillator strength $x_k$
      of the $k^{th}$ dipole mode is $x_k=y_k^2$ where
      \begin{equation}\label{9}
       y_k =  \int r \cos \vartheta \,w_k(\mbox{\boldmath $r$})
      \,d\mbox{\boldmath $r$} \, .
      \end{equation}
      I recall that the functions $s_k(\mbox{\boldmath $r$)}$ and
      $w_k(\mbox{\boldmath $r$)}$ are normalized
      as indicated previously.
      \par
      The results of computations  are collected
      in Table~1. One may see that the
      contribution of the dipole strength within the range
      7-32 eV to the total figure of 240 electrons required
      by the dipole sum rule is quite large. After the inspection
      of other solutions below this range and above up to 42 eV
      it appear that
      the sum rule is satisfied in defect
      within about then percent. So it is very probably satisfied
      exactly taking into account larger frequencies and
      the computing errors as well.
      The contribution of the modes listed to the static
      polarizability of a free $C_{60}$ give a somewhat large figure,
      in agreement with the quantum-chemical result
      of reference \cite{Fow90}.
      The imaginary part of the dynamic polarizability for
      a free fullerene molecule was recently determined indirectly
      from photoionisation data on gaseous $C_{60}$ \cite{Her92}.
      The authors concluded for the presence of absorption
      peaks at about 7.8, 10, 13, 17 and 21 eV.
      The plasmon collective excitation modes are presumably
      strongly coupled to the single electron states and
      so the absorption line width can be very large.
      Referring to the middle column of the Table~1 we could
      thus suppose that the experimental maxima
      near 13 and 17 eV are due to the two couples of resonances
      here found at 12.14, 14.37 eV and at 16.41, 18.17 eV
      respectively. We may then conclude that the
      prediction of our calculations in that of the
      three cases for which the fullerene radius $r_b$ is the
      nearest to the indicated value of 3.5 \AA \cite{Fisbo}
      is in a good accordance with the experiments.
      I stress however that the aim was especially
      to present here the model and not to present very
      refined predictions.
      The conclusion above still seems to indicate that the method
      outlined does work well. It can be useful to look at
      other multipoles than the simple dipole and then
      investigate such things like the inelastic electron
      scattering and the behaviour of Van der Waals interaction
      at large distance to which large multipoles may contribute
      appreciably \cite{Ash91,Lam92}.
      As a concluding remark I observe that of course the present
      method can be improved further. First such terms like
      exchange and correlation energies must be included in the
      Lagrangian of departure. And on the mathematical side it
      would be useful to correlate the shells thickness
      to the rapidity of variation of $\rho_0^{(i)}$.
\vskip .4 true cm
\noindent {\bf Acknowledgments} \vskip .1 true cm
I like to thank L. Reatto who called my attention to
this subject.
\newpage
%
%             ========    references  =======
%
\setlength{\baselineskip}{22pt}

%
%              ===========     table   =============
%
\newpage
\begin{table}
\caption{The frequencies $\omega$ and the oscillator strengths $x$ for
three values of $r_b$ as result from the present calculation in the
frequency range 7-32 eV. The cases where $x<0.25$ are omitted. \hfil}
\label{table1}
\begin {tabular} {rrrrrrrr} \\ \hline \hline \\
 \multicolumn{2} {c} {$r_b=6.4$}
& & \multicolumn{2} {c} {$r_b=6.6$}
& & \multicolumn{2} {c} {$r_b=6.8$} \\ \\
\multicolumn{1} {c} {$\omega$ (eV)} & \multicolumn{1} {c} {$x$}
& & \multicolumn{1} {c} { $\omega$ (eV)} & \multicolumn{1} {c} {$x$}
& & \multicolumn{1} {c} { $\omega$ (eV)} & \multicolumn{1} {c} {$x$} \\
\hline \\
7.23 &  9.76 & & 7.60 & 11.26 & & 8.00 & 13.11 \\
9.44 & 10.44 && 9.92 & 12.46 && 10.43 & 15.25 \\
11.57 & 14.34 && 12.14 & 18.00 && 12.71 & 21.98 \\
13.70 & 18.03 && 13.85 & 1.41 && 12.95 & 1.39 \\
14.89 & 0.69 && 14.37 & 22.32 && 14.99 & 33.62 \\
15.77 & 25.19 && 16.41 & 36.73 && 16.88 & 37.40 \\
17.71 & 35.55 && 18.17 & 29.75 && 17.61 & 5.79 \\
19.38 & 28.30 && 18.92 & 7.76 && 18.95 & 32.81 \\
20.28 & 4.66 && 20.20 & 27.08 &&  20.85 & 7.51 \\
21.40 & 23.54 && 22.09 & 6.36 && 22.24 & 5.78 \\
23.20 & 6.84 && 23.76 & 2.82 && 23.21 & 4.59 \\
25.82 & 7.83 && 24.42 & 6.32 && 26.87 & 4.94 \\
27.39 & 1.00 && 26.46 & 0.36 && 27.88 & 0.37 \\
30.94 & 4.38 && 28.50 & 1.50 && 30.15 & 0.25 \\
31.74 & 0.58 & & 29.06 & 3.60 & & 31.51 & 4.18  \\
\hline \hline
\end{tabular}
\end{table}
\end{document}